\documentclass[12pt,a4paper]{article}

\usepackage{graphicx}
\usepackage{amsmath}
\usepackage{epsf}
\usepackage{bm}

\usepackage{mathrsfs}
\usepackage{amssymb}
\usepackage{epsfig}

\input epsf.sty

\newcommand{\be}{\begin{equation}}
\newcommand{\ee}{\end{equation}}
\newcommand{\bea}{\begin{eqnarray}}
\newcommand{\eea}{\end{eqnarray}}
\newcommand{\nn}{\nonumber \\}

\begin{document}

\author{\textsc{I. Lyberg}\\
{\normalsize School of Theoretical Physics, Dublin Institute for Advanced Studies}\\
{\normalsize ilyberg@stp.dias.ie}}

\title{A ``quantum spherical model'' with transverse magnetic field.} 

\maketitle

\section{Introduction}
\label{intro}
The Quantum Ising Model with a transverse magnetic field is well known \cite{lse} \cite{mattis}. In one dimension it has the Hamiltonian
\begin{equation} {\cal H}_N = -J \sum_{n=1}^N \sigma^x_n
\sigma_{n+1}^x + B \sum_{n=1}^N \sigma^z_n+H\sum_{n=1}^N \sigma^x_n,
\label{hamiltonian}
\end{equation}
where $J>0$ is the coupling constant and $\sigma^x = \left(
\begin{array}{cc}
0 & 1 \\
1 & 0 
\end{array} \right)$, $\sigma^y = \left(
\begin{array}{cc}
0 & -i \\
i & 0 
\end{array} \right)$
and $\sigma^z = \left(
\begin{array}{cc}
1 & 0 \\
0 & -1 
\end{array} \right)$ are the Pauli matrices. $B$ and $H$ are transverse and longitudinal magnetic fields, respectively. The partition function is 
\be Z_N={\rm tr}~e^{-\beta {\cal H}_N}
\ee
where $\beta$ is the inverse temperature. In the case where $H=0$ this model has been exactly solved \cite{lse} \cite{mccoy}. The free energy is \cite{mattis}
\bea f(\beta,J,B)&=&-\lim_{N\to \infty}\frac{1}{\beta N}\log{Z_N}\nn
                 &=&-\frac{1}{2\pi\beta}\int_{0}^{2\pi}\log{2\cosh{\beta\Delta(x)}}~dx
\eea
where 
\be \Delta(x)=\sqrt{J^2+B^2-2BJ\cos{x}}.
\ee
In particular the ground state energy is given by
\be f_{\infty}(J,B)=\lim_{\beta\to \infty}f(\beta,J,B)=-\frac{1}{2\pi}\int_{0}^{2\pi}\Delta(x)~dx.
\label{gse}
\ee
In this limit there is a critical point in $B$ at $B=J$. The correlation function
\be \langle \sigma^x_j\sigma^x_k\rangle=\lim_{N\to\infty}\frac{{\rm tr}~\sigma^x_j\sigma^x_ke^{-\beta {\cal H}_N}}{Z_N}
\ee
can be written as a Toeplitz determinant of size $|j-k|$ just as the correlation function of the two dimensional classical Ising model \cite{onsager}, but only in the limit $\beta\to\infty$. In fact the correlation function $\lim_{\beta\to\infty}\langle \sigma^x_j\sigma^x_k\rangle$ is the same as the diagonal correlation function $\langle \sigma_{jj}\sigma_{kk}\rangle$ of the two dimensioanl classical Ising lattice for $T<T_c$, the ratio $B/J$ in the one dimensional quantum model corresponding to $(\sinh{2E_1/k_{\rm B}T}\sinh{2E_2/k_{\rm B}T})^{-1}$ in the two dimensional classical model. (Here $E_1$ and $E_2$ are the coupling constants in the horizontal and vertical directions, respectively). In particular the limit of infinite separation is given by \cite{mccoy}
\begin{equation}
\lim_{|j-k|\to \infty}\lim_{\beta\to\infty}\langle \sigma^x_j\sigma^x_k \rangle=
\begin{cases}
\left\{1-(B/J)^2\right\}^{1/4} & \text{if }  B<J,\\
0 & \text{if } B\geq J,
\end{cases}
\label{corr}
\end{equation}
which is most easily proved using Szeg\"o's theorem \cite{mpw} \cite{grenander}.

\section{The quantum spherical model}
In analogy with (\ref{hamiltonian}) we define a partition function of a ($d$-dimensional) isotropic quantum spherical model on a lattice $\Lambda$ as follows:
\bea Z_N&=&\int_{[0,\infty)^N}\int_{[0,2\pi)^N}\int_{[0,\pi]^N}e^{\sum_{j,k\in\Lambda~:~\langle jk\rangle}\beta Jr_j\cos{\theta_j}r_{k}\cos{\theta_{k}}}\nn
&&e^{\sum_{j\in \Lambda}\beta(Br_j\sin{\theta_j}\cos{\varphi_j}+Hr_j\cos{\theta_j})}\nn
&&\prod_{l=1}^Nr_l^2\sin{\theta_l}~d^N\theta~d^N\varphi~\delta\bigg(\sum_{m=1}^Nr_m^2-N\bigg)d^Nr\nn
&=&\int_{{\bf R}^{3N}}e^{\sum_{\langle jk\rangle}\beta Jz_jz_{k}+\sum_j\beta (Bx_j+Hz_j)}\delta\bigg(\sum_{k=1}^N(x_k^2+y_k^2+z_k^2)-N\bigg)d^{3N}{\bf x}.
\label{qsm}
\eea
Here $J>0$, $B\geq 0$ and $H>0$. $\delta$ signifies the Dirac distribution. The notation $\langle jk\rangle$ means that $j$ and $k$ are nearest neighbors on $\Lambda$. Unlike the Quantum Ising Model with $H=0$, in this model the critical point is $B=2Jd$ (in the limit $H\to 0$). In fact, it will be shown that in this limit the ground state free energy $f_{H,\infty}:=-\lim_{\beta \to\infty}\lim_{N \to\infty}(N\beta)^{-1}\log{Z_N}$ is given by
\begin{equation}
f_{0,\infty}=\lim_{H\to 0}f_{H,\infty}=-
\begin{cases}
Jd+B^2/4Jd & \text{if } B\leq 2Jd,\\
B & \text{if } B> 2Jd.
\end{cases}
\label{feq}
\end{equation}
We shall now give a proof of (\ref{feq}).

\subsection{The case $B>2Jd$}
We use the method of steepest descent to prove this result, following the calculation by Baxter \cite{baxter}. We let $H=0$ in (\ref{qsm}). Clearly the integrand in (\ref{qsm}) may be multiplied by a factor $\exp{a(\sum_{k=1}^N(x_k^2+y_k^2+z_k^2)-N)}$ without changing the partition function $Z_N$. Using the identity
\be \delta(x)=\frac{1}{2\pi}\int_{-\infty}^{\infty}e^{isx}ds,
\label{delta2}
\ee
together with (\ref{qsm}) and letting $a>0$, we get
\bea Z_N=\frac{\pi^{N-1}}{2}\int_{{\bf R}^{N}}\int_{-\infty}^{\infty}\Big(\frac{1}{a+is}\Big)^{N}\exp{\frac{N(\beta B)^2}{4(a+is)}}\nn
\exp{[\sum_{\langle jk\rangle}\beta Jz_jz_{k}+\sum_j(a+is)(1-z_j^2)]}~ds~d^{N}z
\label{qsm2}
\eea
after integrating over ${\bf x}$ and ${\bf y}$. Let ${\bf V}$ be the symmetric matrix such that
\be {\bf z}^T{\bf V}{\bf z}=(a+is)\sum_{j=1}^Nz_j^2-\beta J\sum_{\langle jk\rangle}^Nz_jz_{k}.
\label{v}
\ee
In this way (\ref{qsm2}) can be written as 
\bea Z_N=\frac{\pi^{N-1}}{2}\int_{{\bf R}^{N}}\int_{-\infty}^{\infty}\Big(\frac{1}{a+is}\Big)^{N}\exp{\frac{N(\beta B)^2}{4(a+is)}}\nn
\exp{[-{\bf z}^T{\bf V}{\bf z}+N(a+is)]}~ds~d^{N}z.
\label{qsm3}
\eea
We now choose the constant $a$ so large that all the eigenvalues of ${\bf V}$ have positive real part. This allows us to change the order of integration, and we may now write (\ref{qsm3}) as
\bea Z_N=\frac{\pi^{3N/2-1}}{2}\int_{-\infty}^{\infty}\Big(\frac{1}{a+is}\Big)^{N}(\det{{\bf V}})^{-1/2}\nn
\exp{\bigg[\frac{N(\beta B)^2}{4(a+is)}+N(a+is)\bigg]}~ds.
\label{qsm4}
\eea 
We need to calculate the eigenvalues of ${\bf V}$. Since ${\bf V}$ is cyclic, this is easily done. We let the lattice be $d$-dimensional hypercubic, so that 
\be N=L^d
\label{nld}
\ee
for some positive integer $L$. It now follows from (\ref{v}) that the eigenvalues are 
\be \lambda(\omega_1,...,\omega_d)=a+is-\beta J\sum_{j=1}^d\cos{\omega_j}
\label{ev}
\ee 
where each $\omega_j$ takes the values $\{2\pi k/L\}_{k=0}^{L-1}$, and $a>\beta J d$.
The determinant of ${\bf V}$ is the product of its eigenvalues, so
\be \log{\det{{\bf V}}}=\sum_{\omega_j~:~1\leq j\leq d}\log{\lambda(\omega_1,...,\omega_d)}.
\label{logdet}
\ee
Clearly
\be Z_N=\frac{\beta J}{2\pi i}\left(\frac{\pi}{\beta J}\right)^{3N/2}\int_{c-i\infty}^{c+i\infty}e^{N\phi(w)}dw,
\label{z}
\ee
where 
\be \phi(w)=\beta Jw-\frac{1}{2}g(w)+(\beta B)^2/4\beta Jw,
\label{phi2}
\ee
$c=(a-\beta Jd)/\beta J$ and
\be g(z)=2\log{w}+\frac{1}{N}\sum_{\omega_j}\log{(w-\sum_{j}\cos{\omega_j})}.
\ee
Since $\phi$ approaches $+\infty$ as $w$ approaches 0 or $+\infty$ along the real line, $\phi$ has a minumum 
at some $w_0$, $0<w_0<\infty$. Thus $\Re{\phi}$ has a maximum at $w_0$ along the line $(w_0-i\infty,w_0+i\infty)$.
Since $B>2Jd$, we may choose $c=w_0$. We now use the method of steepest descent (see for 
instance Murray \cite{murray}), by letting $N$ approach infinity. In this way, the free energy is
\bea f&=&-\beta^{-1}\lim_{N\to \infty}N^{-1}\log{Z_N}\nn
       &=&-\frac{3}{2\beta}\ln{(\pi/\beta J)}-\beta^{-1}\phi(w_0). 
\eea
Now 
\be \lim_{\beta\to\infty}w_0=B/2J,
\ee
and thus the ground state energy is
\bea \lim_{\beta\to\infty}f & = & -\lim_{\beta\to\infty}\beta^{-1}\phi(w_0)
\nn
& = & -B.
\eea

\subsection{The case $B\leq 2Jd$}
In this case we let $H>0$, so instead of (\ref{qsm3}) we have
\bea Z_N=\frac{\pi^{N/2-1}}{2}\int_{{\bf R}^{N}}\int_{-\infty}^{\infty}\Big(\frac{1}{a+is}\Big)^{N}\exp{\frac{N(\beta B)^2}{4(a+is)}}\nn
\exp{[-{\bf z}^T{\bf V}{\bf z}+{\bf h}^T{\bf z}+N(a+is)]}~ds~d^{N}z,
\label{qsm3b}
\eea
where ${\bf h}=\beta H (1,...,1)$. We change variables to ${\bf t}={\bf z}-\frac{1}{2}{\bf V}^{-1}{\bf h}$, and 
rotate the axes in $(t_1,...,t_N)$ to make ${\bf V}$ diagonal. Thus we get
\bea Z_N=\frac{\pi^{N/2-1}}{2}\int_{-\infty}^{\infty}\Big(\frac{1}{a+is}\Big)^{N}(\det{{\bf V}})^{-1/2}
\exp{\frac{N(\beta B)^2}{4(a+is)}}\nn
\exp{[{\bf h}^T{\bf V}^{-1}{\bf h}/4+N(a+is)]}~ds.
\label{qsm4b}
\eea
Thus
\bea Z_N=\frac{\pi^{N/2}}{2\pi i}\int_{c-i\infty}^{c+i\infty}
e^{N\phi(w)}~dw,
\eea
where $a+is-\beta Jd=\beta Jw$ and
\bea \phi(w) & = & \beta J (w+d)+\frac{(\beta H)^2}{4\beta J w}+\frac{(\beta B)^2}{4\beta J(w+d)}
\nn
& - & \log{\beta J(w+d)}
-\frac{1}{2}\sum_{\omega_j}\log{(\beta J(w+d)-\beta J\sum_{j}\cos{\omega_j})}.
\eea
We proceed in the same way as before, taking the limit $N\to\infty$ and then $\beta\to \infty$. In this case 
$w_0\to 0$ as $H\to 0$. The free energy is thus 
\be f=-Jd-\frac{B^2}{4Jd}.
\ee
This ends the proof.

\section{Discussion}
Comparison of (\ref{gse}) and (\ref{feq}) shows that the susceptibilites of the two models at $B=0$ are equal 
when $d=1$; 
that is $-\partial^2f_{\infty}/\partial B^2|_{B=0}=1/2J$ and $-\partial^2f_{0,\infty}/\partial B^2|_{B=0}=1/2Jd$. 
While the Quantum Ising Model has only been exactly solved in the one dimensional case, 
the quantum spherical modelcan be solved in any finite dimension. 
\\
\\
\textbf{Acknowledgements}
The authour would like to thank Prof. T. Dorlas for many discussions.

\end{document}